# Determining Curie temperature of (Ga,Mn)As samples based on electrical transport measurements: low Curie temperature case.

Adam Kwiatkowski,[1, a)] Marta Gryglas-Borysiewicz,[1] Piotr Juszyński,[1] Jacek Przybytek,[1] Maciej Sawicki,[2] Janusz Sadowski,[2, 3] Dariusz Wasik,[1] and Michał Baj[1]

[1)] *Institute of Experimental Physics, Faculty of Physics, University of Warsaw, Pasteura 5, 02-093 Warsaw, Poland*

[2)] *Institute of Physics, Polish Academy of Sciences, al. Lotników 32/46, 02-668 Warsaw, Poland*

[3)] *MAX-IV laboratory, Lund University, P.O. Box 118, 221 00 Lund, Sweden*

(Dated: 9 May 2016)

In this paper we show that the widely accepted method of the determination of Curie temperature ($T_C$) in (Ga,Mn)As samples, based on the position of the peak in the temperature derivative of the resistivity, completely fails in the case of non-metallic and low-$T_C$ unannealed samples. In this case we propose an alternative method, also based on electric transport measurements, which exploits temperature dependence of the second derivative of the resistivity upon magnetic field.

Throughout the last two decades of extensive research, (Ga,Mn)As dilute ferromagnetic semiconductor has become a reference model material in which ferromagnetism comes from the coupling between local magnetic moments and delocalised holes.[1] Research in this area is mainly carried out on properly annealed (Ga,Mn)As samples with high Curie temperatures ($T_C$) from 80 K to 185 K (see e.g. Ref.[2,3]). In order to determine $T_C$ one usually uses sensitive SQUID magnetometers,[2] however, very often measurements of the electrical transport properties are used instead, because of their simplicity. One of the basic methods of $T_C$ determination relies on the fact that for such a samples (similar to ferromagnetic metals[4,5]) temperature dependence of the derivative of the zero magnetic field resistivity with respect to the temperature, $d\rho_{xx}/dT$, exhibits an asymmetric peak whose maximum is close to the $T_C$ established from magnetometry[2,3]. The physical model of this phenomenon assumes scattering of charge carriers on critical magnetization fluctuations[4,6] and leads to description of $d\rho_{xx}(T)/dT$ by the function[4]

$$\frac{d\rho_{xx}}{dT}(T) = \begin{cases} \frac{A_1}{\alpha_1}\left(\left|1-\frac{T}{T_C}\right|^{-\alpha_1}-1\right)+B_1, & T \geq T_C \\ \frac{A_2}{\alpha_2}\left(\left|1-\frac{T}{T_C}\right|^{-\alpha_2}-1\right)+B_2, & T < T_C \end{cases} \quad (1)$$

Here, $\alpha_1$ and $\alpha_2$ are critical exponents, $A_1$, $B_1$, $A_2$ and $B_2$ are the coefficients determined from the fit of the function outside the critical area. At $T=T_C$ Eq. (1) predicts a singularity for $\alpha_{1,2} > 0$ or a sharp not differentiable peak for $\alpha_{1,2} \in (-1,0)$. However, inhomogeneities present in real samples should smear any of such sharp structures. This smearing is seen in temperature dependence of magnetization and was reasonably described by the convolution of the unbroadened dependence with an assumed Gaussian distribution of $T_C$ in the sample, centered at an average Curie temperature, $T_{CM}$.[2,7] Although $\rho_{xx}$ is not an extensive physical property, the same method of broadening was successfully applied for the analysis of $d\rho_{xx}/dT$ of annealed, relatively high $T_C$ (Ga,Mn)As samples.[2,3] Here the maximum of $d\rho_{xx}/dT$, due to asymmetry of the structure described by Eq. (1), is not only broadened but also shifted from $T_{CM}$ towards lower temperatures, however while determining the value of $T_{CM}$ this shift can be easily accounted for. All of the relatively high $T_C$ (Ga,Mn)As samples reported in the papers by Wang *et al.*[2] and Novák *et al.*[3] exhibited metallic conductivity character, i.e. their resistance did not increase upon lowering temperature below $T_C$. We will show that this was crucial for the success of the applied method of $T_C$ determination and that this method completely fails in the case of non-metallic and low-$T_C$ unannealed samples.

In this paper we present the results of experiments performed on three (Ga,Mn)As samples (the details are given in Table I). They have been grown on semi-

TABLE I. Basic information about the investigated layers of (Ga,Mn)As and a summary of the experimental results.

| Sample name | A984 | A962 | MA4 |
|---|---|---|---|
| Mn content[a] (%) | 7 | 6 | 2 |
| layer thickness (nm) | 10 | 50 | 50 |
| buffer type | $In_{0.3}Ga_{0.7}As$ | HT GaAs | HT GaAs |
| buffer thickness (nm) | 800 | 100 | 100 |
| annealing | 3 h at 220 °C | none | none |
| strain[b] $\epsilon_{xx}$ (%) | +2 | −0.18 | −0.087 |
| Magnetic anisotropy | out-of-plane | in-plane | in-plane |
| $T_C$[c] (K) | 105 | 81 | 15 |
| T at max. $d\rho_{xx}/dT$ (K) | 91.3 ± 0.2 | 78.4 ± 0.1 | - |
| T at max. $\chi_i^2$ (K) | n.a. | 82.1 ± 0.2 | 16.1 ± 0.1 |

[a] Determined during the growth from the measurement of the difference between growth rates of LT GaAs and (Ga,Mn)As via RHEED intensity oscillations[8,9] with the precision of ±0.1%.
[b] Determined from X-ray diffraction experiments.
[c] Obtained from SQUID magnetometry.

[a)]Electronic mail: adam.kwiatkowski@fuw.edu.pl

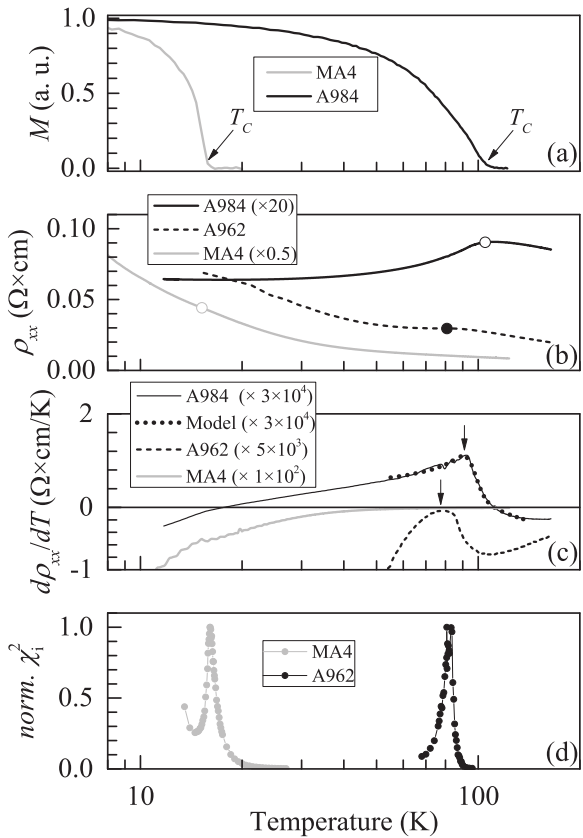

FIG. 1. (a) Temperature dependence of the projection of the spontaneous magnetization on the easy direction for samples MA4 and A984. (b) The dependence of the resistivity of the three measured samples on temperature. (c) The temperature derivative of the curves from Fig. 1.(b) The dotted line is a model function fitted to the experimental result obtained on sample A984. (d) Normalized square of initial susceptibility as a function of temperature derived indirectly from the second derivative of the resistivity with respect to the magnetic field.

insulating GaAs (001) substrates by low-temperature molecular beam epitaxy (MBE). Prior to (Ga,Mn)As deposition, first the substrates have been covered by either $In_{0.3}Ga_{0.7}As$ (sample A984) or High Temperature (HT) GaAs main buffer layer and afterwards an additional thin (about 5 nm) highly resistive Low Temperature (LT) GaAs one. In the case of sample A984 InGaAs buffer has been used to intentionally induce an out-of-plane magnetic anisotropy. After the growth this sample has been annealed to remove interstitial manganese. In our samples the content of manganese atoms located at Ga sites in the GaAs host lattice is ranging from 2% to 7%.

Figure 1.(a) shows temperature dependence of the projection of spontaneous magnetization on the easy direction for sample MA4 with the smallest content of manganese and in-plane anisotropy, as well as for sample A984 with higher Mn content and out-of-plane anisotropy. One can see that these (Ga,Mn)As layers exhibit pure dilute ferromagnetic characteristics[1] with low Curie temperature. All the three layers have Curie temperatures, found from SQUID magnetometry, from 15 K to 105 K (see Table I).

To determine the electric transport properties the Hall-bars oriented along [110] of (Ga,Mn)As have been prepared. Figure 1.(b) presents temperature dependence of the resistivity. Sample A984 is metallic with a clear maximum of $\rho_{xx}$ in the vicinity of $T_C$. The two other samples show insulating character with a pronounced rise of $\rho_{xx}$ at low temperatures. In the case of A962 the maximum of $\rho_{xx}(T)$ is still resolved, while for MA4 it is not observed at all. Thus, there is a qualitative difference between metallic samples (e.g. A984 or other high-$T_C$ (Ga,Mn)As samples[2,3]) and the insulating ones. These clear differences can be seen even better in Fig. 1.(c) presenting temperature derivatives of the experimental curves shown in Fig. 1.(b). For sample A984 the shape of $d\rho_{xx}/dT$ is typical for this class of samples,[2,3] while for MA4 it is completely different. Although for sample A962 there is also a maximum of $d\rho_{xx}/dT$, its shape significantly differs from the one obtained for sample A984. The values of temperature at the maxima of $d\rho_{xx}/dT$ are given in Table I.

Before further analysis of the results we need to discuss possible parasitic effects due to the substrate and buffers. At temperatures 200 - 300 K GaAs substrates exhibit activation type of resistivity with an activation energy exceeding 600 meV. At low temperatures resistivity is immeasurably high. Hence despite of its significant thickness in comparison with (Ga,Mn)As layer the contribution of the substrate to electrical transport below 300 K is negligible. Separate discussion requires sample A984 with relatively thick InGaAs buffer. Here $\rho_{xx}(T)$ points on the metallic character of this sample, thus possible contribution from conductivity of buffer can be excluded in this case. Indeed, taking into account the typical background impurity level in the non-doped, MBE grown layers (at most $10^{15}$ cm$^{-3}$), and rather low expected carrier mobility in relaxed InGaAs buffer (due to high density of misfit dislocations) one gets at least two orders of magnitude higher resistance of 1 $\mu$m thick buffer layer in comparison to that of 10 nm thick, highly doped metallic (Ga,Mn)As one. Thus we conclude, that the transport data presented here reflect solely the (Ga,Mn)As properties.

As was already described above, for high-$T_C$ samples the analysis of the $d\rho_{xx}/dT$ relies on the fitting of Eq. (1) convoluted with Gaussian to the experimental data. Such a procedure has been performed for sample A984. We have assumed that the critical exponents $\alpha_1$ and $\alpha_2$ are both equal to $-0.12$ as predicted by the three-dimensional Heisenberg model.[10] We have checked that their values are not essential for good quality of the fit, which results from quite substantial broadening of the critical structure (the standard deviation of the Gaussian distribution of $T_C$ is $\sigma_{TC} = 4.8$ K). This means that, our experimental results obtained on sample A984 cannot distinguish between various possible models of ferromagnetism. We have obtained $T_{CM} = 96.3$ K. $A_1$ and $A_2$ are very similar: $0.74 \times 10^{-5}$ $\Omega\times$cm/K and $0.84 \times 10^{-5}$ $\Omega\times$cm/K, respectively. An asymmetric shape of the $d\rho_{xx}/dT$ curve in the vicinity of $T_C$ comes mainly from different values of $B_1$ and $B_2$: $-1.25 \times 10^{-5}$ $\Omega\times$cm/K and $0.92 \times 10^{-5}$ $\Omega\times$cm/K, respectively. According to the model, $T_{CM}$ for this sample is 5 K greater than the position of the maximum in $d\rho_{xx}/dT$ (see Table I). This difference is very close to the value of $\sigma_{TC}$. The result of the fitting is presented in Fig. 1.(c) by the dotted line. One can see that the model describes the result of the experiment reasonably well. The same applies to the results of similar fittings presented in the paper by Wang et al.[2] for high-$T_C$ samples. However, for sample A962, and especially for MA4 one it is not possible to explain experimental $d\rho_{xx}/dT$ dependences using the same type of a model. From the inspection of Figures 1.(b) and 1.(c) it is clear that this is mainly due to resistivity background (related to nonmetallic character of the sample) which is not constant in the vicinity of the $T_C$, but it is rapidly increasing with decreasing temperature, what is not reflected by the model (Eq. (1)), where $B_1$ and $B_2$ are assumed to be constant. In addition, in the case of sample MA4 the background is much larger in magnitude than the component associated with the critical scattering. A closer inspection of $d\rho_{xx}/dT$ dependence for this sample shows that in the vicinity of 15 K $d\rho_{xx}/dT$ somehow changes its trend and this temperature is close to $T_C$ evaluated from the magnetometry. This may be in fact the trace of critical scattering in the sample, but it is not reliable to determine $T_C$ value from this data. To visualize the effect of this resistivity background on the shape of the $d\rho_{xx}(T)/dT$ dependence we have used very simple modeling. Namely, beside the critical scattering contribution to resistivity, we have added another one – thermally activated background related to the insulating character of the sample: $\rho_b = \rho_0 \exp\left(\frac{E_a}{k_B T}\right)$, where $E_a$ is an activation energy. Such background is qualitatively consistent with the low-temperature part of $\rho_{xx}(T)$ dependences experimentally observed for samples A962 and MA4 (Fig. 1.(b)). Taking into account these two contributions we end up with the equation describing the temperature dependence of the total resistivity of the sample:

$$\frac{d\rho_{xx}^{tot}}{dT}(T) = -\frac{\rho_0 E_a}{k_B T^2} \times \exp\left(\frac{E_a}{k_B T}\right) + \\ + \int \frac{d\rho_{xx}}{dT}(T, T'_C) \times g(T'_C, T_{CM}) dT'_C \quad (2)$$

where $d\rho_{xx}(T, T'_c)/dT$ is given by Eq. (1) with substitution of $T_C$ by $T'_C$ and $g(T'_C, T_{CM})$ stands for Gaussian distribution of Curie temperatures in the sample, with some standard deviation $\sigma_{TC}$ and mean Curie temperature $T_{CM}$. Basing on Eq. (2) we have qualitatively modeled $d\rho_{xx}(T)/dT$ curves observed for samples A962 and MA4 (Fig. 1.(c)). The second term in the equation was calculated using the same values of parameters as given above for sample A984 except for $T_{CM}$ which has been shifted to 50 K, to approach $T_C$ of these two samples. Concerning the first term in Eq. (2), we have taken constant value of the activation energy $E_a = 1$ meV. This seems to be a reasonable choice, since the change of $\rho_{xx}$ for sample MA4 between 10 K and 100 K is close to 700% (Fig. 1.(b)), which corresponds to $E_a = 1.8$ meV and for sample A962 it should be smaller. To parametrize the relative contribution of the first term in Eq. (2) in comparison to the second one, we have used the dimensionless coefficient $\kappa = \frac{\rho_0 E_a |\alpha_2|}{A_2 k_B T_{CM}^2}$ ($A_1$ and $A_2$ are practically the same and $\alpha_1 = \alpha_2$, so it does not matter whether we use here indexes "1" or "2"). Figure 2 shows the results of calculations performed for a set of values of $\kappa$. The higher the contribution of the thermally activated background to the total resistivity (i.e. higher $\kappa$), the bigger the deviation of $d\rho_{xx}/dT$ from its canonical shape seen for sample A984. Moreover, even if the maximum of $d\rho_{xx}/dT$ is still visible, the magnitude of the background influences its position. It means that the evaluation of $T_C$ from the position of the maximum of $d\rho_{xx}/dT$ leads to systematic errors in following the evolution of $T_C$ upon some external parameters (like e.g. hydrostatic pressure), since these external parameters can also influence the background. This was already noticed in the paper by Gryglas-Borysiewicz et al.[11]. The structure calculated for $\kappa = 0.4$ resembles experimental result obtained for sample A962, while the one calculated for $\kappa = 2.0$ resembles the result obtained for sample MA4 (Fig. 1.(c)). This all confirms that in the case of low-$T_C$ samples the effect of resistivity background makes it really difficult, or even impossible, to determine the Curie temperature in the same way as it can be done in high-$T_C$ case. In this case we propose an alternative method which provides the possibility to observe a really sharp structure related to $T_C$, which evolution upon an external parameter (like pressure) can be followed.

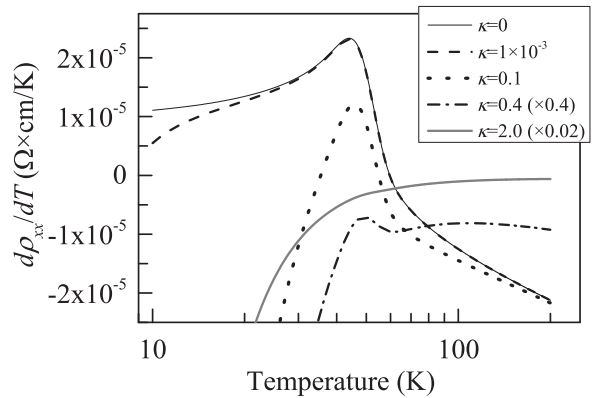

FIG. 2. Modelling of $d\rho_{xx}/dT$ for different values of $\kappa$.

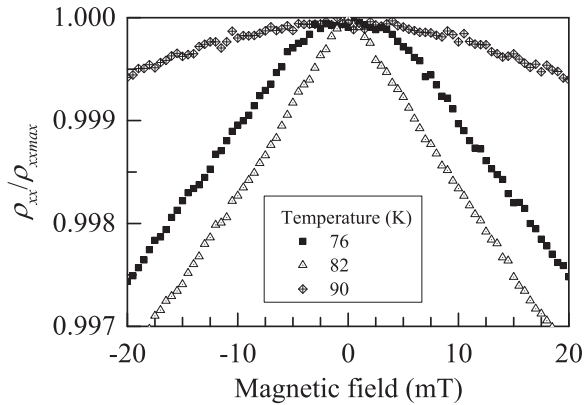

FIG. 3. The dependence of the resistivity on the external magnetic field, measured on sample A962 at three temperatures close to the $T_C$.

It is well known that galvanic properties of magnetic samples probe their magnetic properties. The basic model of the resistivity tensor[12] assumes that the sample is a single magnetic domain with the magnetization $M$. In the case of samples MA4 and A962, with magnetization lying in plane due to anisotropy, $\rho_{xx}$ takes the form:[12]

$$\rho_{xx} \approx \rho_1 + R_1 M^2 \cos^2\theta_M \qquad (3)$$

where $\theta_M$ is the angle between current density and magnetization, $\rho_1$ and $R_1$ are coefficients describing the microscopic properties of the material. The measurement of $\rho_{xx}$ enables indirect determination of $M^2$. There is an unsolved problem how to construct analytical models that take into account the existence of the magnetic domain structure. Following Haigh et al.[13] we assume that Eq. (3) can be applied to real samples, if we only use their average macroscopic magnetization. If we apply external magnetic field $B$ of a very small amplitude in the plane of the layer, which does not cause irreversible changes in the sample magnetization, the second derivative of $\rho_{xx}$ with respect to $B$ will be proportional to the square of the initial susceptibility $\chi_i$ of the sample:

$$\left.\frac{d^2\rho_{xx}}{dB^2}\right|_{B=0} \propto \chi_i^2 \qquad (4)$$

Figure 3 presents resistivity as a function of external magnetic field obtained for sample A962 at three temperatures: $T < T_C$, $T \approx T_C$ and $T > T_C$. It is clear that the absolute value of the second derivative of $\rho_{xx}$ at $B = 0$ has a maximum close to $T_C$. The effect is very noticeable and quite easy to measure. Figure 1.(d) shows $\chi_i^2$ as a function of temperature derived indirectly from $d^2\rho_{xx}(T)/dB^2$ at $B = 0$ for samples A962 and MA4. The result shows sharp, clear peaks. The values of temperatures corresponding to its maxima have been listed in Table I. We associate the presence of these peaks with the Hopkinson effect[14] wherein, the initial magnetic susceptibility increases with increasing temperature and a maximum occurs just below $T_C$. The effect is analogous to the peak observed near the Curie temperature in AC magnetometry of (Ga,Mn)As samples.[15,16] When the magnetic system becomes ferromagnetic, just below $T_C$ the energy density of magnetocrystalline anisotropy rapidly decreases when temperature increases and it is easier to rotate the magnetization of magnetic domains by applying magnetic field of a small amplitude. At the same time, with the increase of temperature also magnetic domain walls are becoming softer. Gradually it is easier to make reversible deformation by applying magnetic field of a small amplitude. This causes that below $T_C$ the initial susceptibility increases with increasing temperature. In the paramagnetic phase, to a first approximation, we expect that susceptibility will decrease with temperature as in Curie - Weiss law. These mechanisms lead to the observation of the structure in the form of a peak in temperature dependence of initial susceptibility. In the case of low-$T_C$ (Ga,Mn)As these characteristic features of the $\chi_i^2$ are much more pronounced than the features associated with the phase transition seen in $d\rho_{xx}/dT$.

To conclude, we want to emphasize that in the case of high Curie temperature (Ga,Mn)As samples, if below $T_C$ the conductivity clearly exhibits metallic character, the temperature corresponding to the maximum of $d\rho_{xx}/dT$ is really near the average value of the Curie temperature. Moreover, the model of the phenomenon described in the work by Wang et al.[2] allows to assess the sample inhomogeneity and the relationship between the temperature corresponding to the maximum of $d\rho_{xx}/dT$ and the average value of the Curie temperature in the macroscopic sample. However, in low-$T_C$ unannealed samples it is quite common that below $T_C$ the conductivity clearly exhibits insulating character and the characteristic maximum in $d\rho_{xx}/dT$ either lies on the background which is changing monotonically with temperature or even does not exist at all. This background makes it difficult or even impossible to determine $T_C$ using this method. We have shown, however, that in such samples one can see very clear structures associated with the phase transition in the temperature dependence of the initial magnetic susceptibility derived indirectly from the measurement of $d^2\rho_{xx}(T)/dB^2$ at $B = 0$. For such samples we have shown the evident advantage of this method of $T_C$ determination over the first one, based on the maximum of $d\rho_{xx}/dT$.

The paper has been supported by the National Science Center (Poland), grant No. DEC-2011/03/B/ST3/03287.